\documentstyle[preprint,aps]{revtex}
\begin{document}
\title{Why are glass-forming liquids non-Arrhenius?}
\draft
\author{Jeppe C. Dyre}
\address{Department of Mathematics and Physics (IMFUFA), 
Roskilde University, P.O.Box 260, DK-4000 Roskilde, Denmark}
\date{\today}
\maketitle{}
\begin{abstract}

A major mystery of glass-forming liquids is the non-Arrhenius
temperature-dependence of the average relaxation time.
This paper briefly reviews the classical phenomenological models
for this phenomenon - the free-volume model and the entropy model
- and critiques against these models.
We then discuss a recent model [Dyre, Olsen, and Christensen,
Phys.
Rev. B 53, 2171 (1996)] according to which the activation
energy for the average relaxation time is determined by the work
done in shoving aside the surrounding liquid to create 
space needed for a flow event.
In this model the non-Arrhenius temperature-dependence is a
consequence of the fact that the instantaneous
(infinite-frequency) 
shear modulus increases upon cooling.
\end{abstract}


\section{Introduction}

Apparently all supercooled liquids are able to form glasses
\cite{kau48,har76,bra85,joh85,jac86,ang88,sch90,ang91,hun93,%
2ang95,moh95,edi96}.
The glass transition takes place when the viscosity 
of the supercooled liquid upon cooling becomes so large that
molecular motion is arrested on the time-scale of the experiment.
The fascination of this phenomenon lies in the fact that
chemi\-cally very different liquids - involving ionic
interactions, van der Waals forces, hydrogen bonds, 
covalent bonds, or even metallic bonds - 
exhibit a number of common properties when 
cooled to become highly viscous \cite{bra85,ang88,sch90,edi96}.
Of particular interest here is the temperature-dependence of the
average relaxation time, $\tau$.
This quantity may be determined, e.g., as the inverse
dielectric, mechanical or specific heat loss peak frequency.  
Alternatively, it may be calculated from the viscosity $\eta$
and the infinite-frequency shear modulus $G_{\infty}$ by means
of Maxwell's expression 

\begin{equation}\label{1}
\tau\ =\ \frac{\eta}{G_{\infty}}\,.
\end{equation}
These definitions do not give exactly identical $\tau$'s,
but the difference is insignificant for the present purposes.
It is widely believed 
\cite{kau48,bra85,ada65,gre81,gol69,bra84,gol72,wil75,don81,%
don82,dyr87,sti88,cha93,moh94,dyr95} that different
$\tau$'s are roughly identical because they
basically measure the rate of ``flow events'':
Most molecular motion in a highly viscous liquid is purely
vibrational around a potential energy minimum.
Only seldom does real motion take place. 
This happens in the form of a sudden rearrangement of molecules,
a process which is unlikely because of the
large potential energy barrier to be overcome
\cite{kau48,gol69,sti88}.
Kauzmann referred to flow events as a ``jumps of
molecular units of flow between different positions of
equilibrium in the liquid's quasicrystalline lattice''
\cite{kau48}.
The molecules involved in a flow event define a
``relaxing unit'' \cite{kau48}, ``cooperatively rearranging
subsystem'' \cite{sch90} or ``cooperatively rearranging
region'' \cite{ada65}, ``quasi-independent unit'' \cite{gol72},
``thermokinetic structure'' \cite{don82}, ``molecular domain''
\cite{sti88},
or ``dynamically correlated domain'' \cite{cha93}.

As the glass transition is approached, the average relaxation
time becomes longer and longer.
For typical cooling rates
$\tau$ is of order $10^3\ $s at $T_g$.
From a general physical/chemical point of view, the
temperature-dependence of $\tau$
is anomalous in the following sense.  
In only very few liquids is $\tau$ Arrhenius
(examples are ${\rm SiO_2}$, ${\rm GeO_2}$, 
${\rm BeF_2}$ or albite (${\rm NaAlSi_3O_8}$)\cite{bra85}).  
Predominantly, $\tau$ is non-Arrhenius by
exhibiting an apparent activation energy
[$\partial\ln\tau/\partial(k_BT)^{-1}$] that increases
as the temperature decreases .
A measure of the departure from non-Arrhenius behavior is the
fragility $m$, defined as the apparent activation energy 
at $T=T_g$ in units of $k_BT_g\ln(10)$ \cite{ang85}.  
For a simple Arrhenius liquid $m$ is about 16; for most viscous
liquids $m$ is between 50 and 150.
Liquids with large fragility are termed ``fragile'', 
liquids with fragility not far above $16$ are termed
``strong'' \cite{ang85}.
There is a general tendency that fragile liquids have broader
distributions of relaxation times than strong liquids
\cite{boh93,boh94}.
This rule, however, is not without exceptions
\cite{joh76,tor90}.

In the discussion below we will not distinguish between the 
temperature-dependence of average relaxation time and of
viscosity, because
these two quantities are roughly proportional 
(in Eq.\ (\ref{1}) the temperature-dependence of $G_\infty$ is 
insignificant).
We identify the temperature-dependent activation energy from the
expression \cite{dyr95,kiv96} 

\begin{equation}\label{1.5}
\tau\ =\ \tau_0\ \exp\left(\frac{\Delta E(T)}{k_BT}\right).
\end{equation}
Although $\Delta E(T)$ is different from the apparent activation
energy,
experiment imply that $\Delta E(T)$ also increases as the
temperature decreases.

It is not at all obvious that a general explanation for
the non-Arrhenius behavior of chemically quite different viscous
liquids exists, but it seems to be a reasonable first hypothesis.
This paper discusses models for the non-Arrhenius average
relaxation time, models that are phenomenological in the sense
that $\tau(T)$ is determined by some macroscopic property of the
liquid.
The most famous phenomenological models for the non-Arrhenius
$\tau$'s are the {\it free-volume model} of Cohen, Turnbull and
Grest \cite{gre81,coh59,tur61,tur70} and the {\it entropy model}
of Gibbs, DiMarzio and Adam \cite{ada65,gib58}.
These models and critiques against them are briefly reviewed
below (see also Johari's review of phenomenological models Ref.
\cite{joh85}).
We then discuss a recently proposed model \cite{dyr96,pisa},
according
to which the activation energy of a flow event mainly 
originates in the work done in shoving aside the surrounding
liquid to
create enough space for the molecules to rearrange.

\section{Early Phenomenological Models}

The importance of volume was stressed long time ago by Eyring and
coworkers, who suggested that the viscosity of a liquid is
lower the greater the number of holes present \cite{ros41}.
Defining the free volume per molecule $v_f$ as the
average volume per molecule in the liquid minus the volume of the
molecule itself, Doolittle \cite{doo51} in 1951 found that the
viscosity of a number of simple hydrocarbon liquids may be fitted
by the expression

\begin{equation}\label{2}
\eta\ =\ \eta_o\ \exp\left({\frac{C}{v_f}}\right)\,.
\end{equation}
In 1959 this expression was derived by Cohen and Turnbull
arguing as follows \cite{coh59}.
The molecules are modelled as hard spheres.
A molecule is mostly confined to a cage bounded by its
immediate neighbors.
Occasionally, there is a fluctuation in density which opens up a
hole within the cage.
Molecular transport occurs only when a void having a volume
greater than some critical value $v^*$ forms.
The total free volume may be distributed in various ways between
the cages.
The average relaxation time is essentially the inverse of the
probability $P$
that redistribution of free volume by chance creates a void of
greater volume than $v^*$.
Turnbull and Cohen calculated this probability by standard
statistical mechanical arguments \cite{coh59}.
Their result is $P\propto\exp(-C/v_f)$, leading to Eq.\ (\ref{2})
via Eq.\ (\ref{1}).
A basic assumption in the free-volume model is that no energy is
required for redistribution of free volume.
When the model is applied to real liquids, Cohen and
Turnbull {\it defined} the free volume as that part of the excess
volume that may be redistributed with no increase in energy
\cite{tur61}. 

In the free-volume model the temperature-dependence of average
relaxation time
comes from the fact that the free volume decreases with
decreasing temperature.
If the free volume is taken to be a linearly decreasing function
of temperature, one arrives at the famous 
Vogel-Fulcher-Tammann (VFT) expression,

\begin{equation}\label{3}
\tau\ =\ \tau_0\ \exp\left({\frac{A}{T-T_0}}\right)\,.
\end{equation}
Here $T_0$ is the temperature at which there is no free volume.

It is noteworthy that in the approach of Cohen and Turnbull, the
concept of free
volume has a meaning different from that of Doolittle
\cite{doo51}.
In Doolittle's definition, the molecular volume is obtained by
extrapolating the
liquid volume to zero temperature and consequently
the free volume is zero only at zero temperature.
Thus, while the experimentally motivated Eq.\ (\ref{2}) was the
starting point
of Cohen and Turnbull, their theory represents a quite different
way of thinking than that of Doolittle.

What critiques may be raised against the free-volume model?
The derivation of Eq.\ (\ref{2}) may be questioned because of the
primitive way in which the entropy of the free volume is taken
into account.
Moreover, despite several attempts \cite{gre81,tur61} the
very concept of free volume in Cohen and Turnbull's sense seems
to be ill
defined operationally for general liquids.
When it comes to a comparison to experiment, the VFT-equation
Eq.\ (\ref{3}) often gives a good fit to data \cite{sch90,ang91}.
However, the fit is seldom perfect; in particular, there are
systematic deviations close to $T_g$, where the 
average relaxation time is
apparently always less temperature-dependent than predicted by
Eq.\ (\ref{3}) \cite{sch90,kiv96,bar66,twe71,lau72,sti95}.
Since Eq.\ (\ref{3}) is derived by combining the free-volume
model with the {\it ad hoc} postulate that the free volume
depends linearly on temperature, a more direct test of the model
may be performed by applying pressure to the liquid.
The model predicts that the average relaxation time
is solely a function of density.
Indeed, $\tau$ does increase dramatically at high
pressures, but quantitatively the free-volume model is not
confirmed \cite{bra85}.
A further test of the free-volume model is based on the fact that
the glass transition is similar to a second-order phase
transition in the sense of Ehrenfest (with continuity of first
derivatives of the free energy and discontinuity of the second
derivatives).
For the pressure-dependence of the transition temperature,
the claim that the average relaxation time is controlled by
volume translates into the requirement that the glass transition
takes place at constant volume.
This implies \cite{dav53}

\begin{equation}\label{4}
\frac{dT_g}{dp}\ =\ 
\frac{\Delta\kappa}{\Delta\alpha}\,,
\end{equation}
where $\Delta\kappa$ is the difference between the isothermal
bulk compressibility of liquid and glass and
$\Delta\alpha$ the same difference for the isobaric thermal
expansion coefficient.
Equation (\ref{4}) is seldom fulfilled \cite{gol63,ang76}.

In the free-volume model the glass transition occurs at a fixed
{\it volume}.
In the theory of Gibbs and DiMarzio from 1958 \cite{gib58} the
variable controlling the average relaxation time is the
configurational {\it entropy}. 
Evaluating the partition function for a lattice model of linear
polymeric chains in a mean-field approximation, Gibbs and
DiMarzio found that there is a
second-order phase transition at a finite temperature $T_K$ 
to a low-temperature state of zero
configurational entropy.
This state is a ``ground state'' of amorphous packing.
Furthermore, Gibbs and DiMarzio argued that in the neighborhood
of $T_K$ the energy barrier restricting transitions between
different molecular configurations is very high,
because ``the few states that could conceivably occur
close to $T_K$ are widely separated in phase space, so proceeding
from one to another involves a considerable change in the
topology of molecular entanglements''.
In this picture, the very equilibrium properties of a supercooled
liquid give rise to kinetic sluggishness which prevents the
equilibrium second order phase transition from being reached in
finite time.

These ideas were quantified in 1965 by Adam and Gibbs
\cite{ada65}.
They argued that the size of cooperatively rearranging
regions, defined as ``the smallest regions that can undergo a
transition to a new configuration without a requisite
simultaneous configurational change on and outside its
boundary'',
diverges as the configurational entropy goes to zero.
The region size is estimated by requiring that at
least two different configurational states should reside in a
region, leading to a size inversely proportional to the
configurational entropy, $S_c$.
If the energy barrier to be overcome is assumed to be
proportional to region size, Adam and Gibbs arrived at the
following
expression for the temperature-dependence of the average
relaxation time,

\begin{equation}\label{5}
\tau\ =\ \tau_o \exp\left({\frac{C}{S_cT}}\right)\,.
\end{equation}
Close to $T_K$ the denominator $S_cT$ may be expanded to first
order
in $T-T_K$ whereby Eq.\ (\ref{5}) becomes the VFT-expression Eq.\
(\ref{3}) with

\begin{equation}\label{6}
T_0\ =\ T_K\,.
\end{equation}

The entropy model resolves the Kauzmann paradox of a negative
configurational entropy below $T_K$ without just avoiding it
(as Kauzmann did himself \cite{kau48} by suggesting that
crystallization sets in before $T_K$ is reached). 
In many respects the model is in good agreement with experiment.
Thus, Eq.\ (\ref{6}) is often obeyed
\cite{ang91,ada65,ang76,ang74}, which is quite remarkable, given
the dynamic definition of $T_0$ and the quite different purely
thermodynamic definition of $T_K$.
In particular, systems with only small excess specific heat 
(implying less pronounced Kauzmann paradoxes and $T_K$'s close to
zero), generally tend to be ``strong'', i.e., have VFT $T_0$'s
close to zero  \cite{ang88a}.

The entropy model undoubtedly presents a beautiful scenario.
Still, it may be critiqued both in regard to its relation to
experiment and in regard to its inner consistency.
Experimentally, there is no proof that a second order phase
transition to a state of zero configurational entropy is
underlying the laboratory glass transition.
In many cases a simple two-level system fit excess
entropy data well \cite{ang72};
more generally, the data may be fitted with a model with only few
energy levels \cite{gol72}.
Also, as mentioned above, the VFT-expression Eq.\ (\ref{3}) fails
close to $T_g$, where data are usually less
temperature-dependent.
Finally, it should be mentioned that the identification of
excess entropy with configurational entropy rests on an
assumption that the glass has the same ``fast'' contribution
to the entropy as the crystal at the same temperature.
As pointed out by Goldstein \cite{gol76}, this
assumption is not always realistic because the glass may have
significant contributions to the ``fast'' specific heat from
anharmonic
vibrations and secondary relaxations not present in the crystal.

In regard to the inner consistency of the entropy model, we first
note that the mean-field solution of the lattice polymer model of
Gibbs and DiMarzio is incorrect and that, in fact, the model has
a positive configurational entropy at all positive temperatures
\cite{guj81,dim97}.
Ignoring this objection and accepting the general idea
of a phase transition to a state of zero configurational entropy,
one may reasonably ask \cite{kiv96}:  
What is the nature of the amorphous ground state, the ``ideal
glass state''?
Since this state is unique a simple description of it 
would be expected; however none has been proposed.
The argument of Adam and Gibbs is also not compelling.
They assumed {\it ad hoc} that the energy barrier to be overcome
is proportional to the size of the cooperatively rearranging
region.
Though this may seem  reasonable, it does not have to be correct.
More generally, approaching a zero-entropy state does not in
itself imply a diverging relaxation time.
There is no compelling link between dynamics and
thermodynamics:
In a master equation description of the dynamics
many different possible forms of transition rates - leading to
quite different relaxation behaviors - are consistent with the
same statistical mechanics.

Instead of focussing on {\it volume} or {\it entropy} as the
variable controlling the relaxation of viscous liquids, 
potential {\it energy} may be the relevant variable, as first
suggested by Goldstein \cite{gol69}.
A number of authors have taken this approach 
\cite{bra85,bra84,gol72,dyr87,sti88,cha93,moh94,dyr95,bas87,%
ark96}.
In the simplest energy controlled models the transition state of
a region is
taken to be temperature-independent with potential energy $E_0$
\cite{bra84,gol72,dyr87}, leading to the following expression for
the
activation energy in terms of the average potential energy of one
region $\overline{E}(T)$ is arrived at \cite{dyr95}:

\begin{equation}\label{8}
\Delta E(T)\ =\ E_0 - \overline{E}(T)\,.
\end{equation}
Since the average potential energy decreases with decreasing
temperature, the activation energy increases.
Qualitatively, this is what is seen in experiment.
However, in order to fit data relatively large regions are
needed, implying much broader relaxation time distributions than
observed \cite{dyr95}.
Therefore, the simple picture does not work and more involved
approaches need
to be taken \cite{ark96}.

\section{Shoving model}

The models above discussed all assume that relaxation depends
only on the state of the region involved.
A completely different approach may be taken, where
the relaxation rate depends only on properties of 
the surrounding liquid:
As starting point we take the fact that molecular interactions 
are strongly anharmonic, i.e., with strong short-ranged
repulsions and weak long-ranged attractions;
as shown by Chandler, Weeks and Andersen \cite{cha83}, this ``van
der Waals'' picture explains a number of phenomena in
liquids.
Next, as in the free-volume model
we assume that space is needed for molecules in a viscous
liquid to rearrange.
The idea is that, because of harsh intermolecular repulsion, 
rearrangement at constant region volume is excessively costly, so
it is much easier for the molecules to spend energy on shoving
aside the surrounding liquid.
If the rearranging molecules constitute a sphere which changes
its radius by
$\Delta r$, the energy cost for expanding is $A (\Delta r)^2$
(because the surrounding liquid may be regarded as an elastic
solid on the short time scale of a flow event).
The energy barrier to be overcome inside the sphere is some
function $f(\Delta r)$, that varies strongly with $\Delta r$.
Minimizing the total energy cost leads to  
$2A\Delta r+f'(\Delta r)=0$.
If the ratio between the ``shoving'' work and the ``inner''
barrier to be overcome is denoted by $\lambda$, we find

\begin{equation}\label{9}
\lambda\ =\ 
\frac{A(\Delta r)^2}{f(\Delta r)}\ =\ 
-\frac{1}{2}\frac{d\ln f}{d\ln\Delta r}\,.
\end{equation}
Because of the strong repulsions one expects this logarithmic
derivative to be numerically much larger than one.
Thus, the shoving work gives the dominant contribution to the
energy barrier.
For simplicity we ignore the ``inner'' contribution to the
activation energy.

To calculate the ``shoving'' work we use the fact that during the
flow event
the surrounding liquid behaves as an elastic isotropic
solid with bulk modulus $K_{\infty}$ and shear modulus 
$G_{\infty}$.
These elastic constants are known to be much more
temperature-dependent in viscous liquids than in simple liquids
or solids (crystals or glasses).
Both $K_\infty$ and $G_\infty$ increases as the temperature
decreases.
The work done on the surroundings depends linearly on these
constants, thus
leading to the observed increase in activation energy with
decreasing
temperature.

Actually, it is only the shear modulus that is important.
To show this we refer to the theory of elasticity of
isotropic media \cite{l+l}, assuming that the
activation volume is relatively small:
If $V$ is the volume of
the cooperatively rearranging region it is assumed that
$\Delta V<<V$.
We identify the activation energy with the elastic energy
stored in the surroundings when the volume of the region has
expanded to $V+\Delta V$.
Remember that
elasticity theory \cite{l+l} concerns the relation between the
stress tensor $\sigma_{ij}$ and the strain tensor $u_{ij}$.
The latter is defined by

\begin{equation}\label{gt14}
u_{ij}\ =\ \frac{1}{2}\left( \partial_i u_j+\partial_j u_i
\right)\,,
\end{equation}
where $\partial_i\equiv\partial/\partial x_i$ and
$u_i$ is the i'th component of the elastic displacement
vector ${\bf u}$.
For an isotropic solid the bulk and shear moduli
$K$ and $G$ are defined \cite{l+l} by 

\begin{equation}\label{gt15}
\sigma_{ij}\ =\ Ku_{ll}\delta_{ij}\ +\ 
2G\left( u_{ij}-\frac{1}{3}\delta_{ij}u_{ll}
\right)\,.
\end{equation}
The equation for static equilibrium is

\begin{equation}\label{gt16}
\partial_i\sigma_{ij}\ =\ 0\,.
\end{equation}
Substituting Eq.\ (\ref{gt14}) into Eq.\ (\ref{gt15}) and
subsequently Eq.\ (\ref{gt15}) into Eq.\ (\ref{gt16}) leads to

\begin{equation}\label{gt18}
\left(K+\frac{1}{3}G\right){\bf\nabla}({\bf\nabla\cdot u})
\ +\ G\ \nabla^2{\bf u}\ =\ 0\,.
\end{equation}
For a purely radial displacement
${\bf\nabla\times u}={\bf 0}$ and thus,
via the well-known vector identity 
${\bf\nabla\times(\nabla\times u)}
={\bf\nabla}({\bf\nabla\cdot u})-\nabla^2{\bf u}$,
we have
$\nabla^2{\bf u}={\bf\nabla}({\bf\nabla\cdot u})$.
When this is substituted into Eq.\ (\ref{gt18}) one finds 

\begin{equation}\label{gt19}
{\bf\nabla}({\bf\nabla\cdot u})=\ 0\,,
\end{equation}
implying that ${\bf\nabla\cdot u}=C_1$,
where $C_1$ is a constant.
The displacement (which is radial) is found by solving
${\bf\nabla\cdot u}=r^{-2}\partial_r(r^2u_r)=C_1$,
leading to $u_r=C_2 r^{-2}+C_1 r/3$.
The latter term diverges as $r\rightarrow\infty$ and thus
$C_1=0$.  
In conclusion ${\bf\nabla\cdot u}=0$, i.e., there is no 
compression of the surroundings during a flow event.

If the radius of the region before the expansion is $R$ and the
change of radius is $\Delta R$, we have since
$\Delta R<<R$

\begin{equation}\label{gt20}
u_r\ =\ \Delta R\ \frac{R^2}{r^2}\,\,(r>R)\,.
\end{equation}
The energy density of an elastic solid is \cite{l+l}
$\frac{1}{2}Ku_{ll}^2+
G\left(u_{ij}-\frac{1}{3}\delta_{ij}u_{ll}\right)^2$.
Since $u_{ll}=0$ the energy density is
given by
$Gu_{ij}u_{ij}=G(u_{rr}^2+u_{\phi\phi}^2+u_{\theta\theta}^2)$
(all mixed terms like, e.g., $u_{r\phi}^2$ are zero because the
displacement is purely radial).
When Eq.\ (\ref{gt20}) is used in the definition of the strain
tensor in polar coordinates, we get
for the energy density 
$6G(\Delta R)^2R^4r^{-6}$.
Thus, the elastic energy is given by

\begin{equation}\label{gt21}
\int_R^{\infty}6G(\Delta R)^2R^4r^{-6}\ (4\pi r^2)dr\ =\ 
8\pi G\ (\Delta R)^2\ R\,.
\end{equation}
Substituting $V=4\pi R^3/3$ and $\Delta V=4\pi R^2 \Delta R$ 
into Eq.\ (\ref{gt21}) we find,
introducing the ``characteristic volume'' 

\begin{equation}\label{gt22}
V_c\ =\ \frac{2}{3}\frac{(\Delta V)^2}{V}\,,
\end{equation}
for the activation energy (with $G\equiv G_\infty(T)$)

\begin{equation}\label{gt23}
\Delta E(T)\ =\ G_{\infty}(T)\ V_c\,.
\end{equation}
For the average relaxation time we thus have \cite{dyr96}

\begin{equation}\label{gt24}
\tau\ =\ \tau_0\ 
\exp\left[\frac{G_{\infty}(T)\ V_c}{k_BT}\right]\,.
\end{equation}

Interestingly, extended mode-coupling theory leads to an
expression resembling Eq.\ (\ref{gt24}), except that $G_\infty$
is replaced by the zero-frequency bulk modulus \cite{sjo90}.
The prediction of Eq.\ (\ref{gt24}) was checked against
experiment on a number of organic liquids in Ref. \cite{dyr96}, 
assuming that the characteristic volume is
temperature-independent (strictly speaking, this assumption is
inconsistent with Eq.\ (\ref{9}), but for a strongly anharmonic
potential the temperature-dependence of $V_c$ is negligible
compared to that of $G_\infty$).
In Ref. \cite{dyr96} the following version of the well-known
``Angell plot'' \cite{ang85} was used:
Instead of plotting the logarithm of the viscosity as function of
$T_g/T$, it was plotted as function of $x=G_{\infty}(T)/T$
(normalized to one at $T_g$).
The model predicts that a straight line should result.
In Fig. 1a of Ref. \cite{dyr96} results are shown based on
measurements of the frequency-dependent shear modulus
covering the frequency range 1mHz-50kHz.
Figure 1b of Ref. \cite{dyr96} shows the data of Lamb and
coworkers from 1967 \cite{bar67}, where $G_\infty$ was obtained
by an ultrasonic standing wave technique operating in the MHz
region.
Finally, Fig. 1c of Ref. \cite{dyr96} shows data for two
liquids, where $G_{\infty}$ was obtained by transverse Brillouin
scattering.
Overall, we found good agreement with the model prediction.

It is generally believed that initial stages of glassy
relaxation proceeds with a temperature-independent activation 
energy, which is a characteristic of the frozen structure. 
According to the present model, however, there is a slight
temperature-dependence of the activation energy for glassy
relaxation, deriving from the fact that $G_{\infty}$ is not
completely temperature-independent in the glassy phase.
The fact that $G_{\infty}$ is measurable means that, if the model
is valid, it is possible to monitor the activation
energy for glassy relaxation directly.
In fact, if one assumes Eq.\ (\ref{gt24}) for the 
non-equilibrium structural
relaxation time, the model gives definite predictions for the
rate of relaxation of $G_{\infty}$ itself. 
Note that, because $G_{\infty}$ determines the relaxation rate of

equilibrium liquid as well as of glass, this quantity is a direct
measure
of the fictive temperature.

\section{Discussion}

Early phenomenological models link the non-Arrhenius average
relaxation time of viscous liquids to configurational
{\it entropy}, free {\it volume}, or potential {\it energy}.
Here, an alternative approach to the non-Arrhenius
problem was taken, linking $\tau(T)$ to the high-frequency
elastic shear
modulus.
The starting point of the shoving model is the fact that
intermolecular forces are strongly anharmonic.
Anharmonicity enters the model at three stages in the
argumentation.
First, the strong repulsions imply that it is very costly
for molecules to rearrange at constant volume (a qualitative
argument reminiscent of the free-volume model).
Secondly, anharmonicity implies that
the shoving work much exceeds
the ``inner'' energy barrier (Eq.\ (\ref{9}), a quantitative
argument).
Finally, the fact that $G_\infty$ depends on temperature is
itself a consequence of anharmonicity.
Note that, according to the model, one expects a liquid to be
more fragile the more anharmonic it is.
This, in fact, is what Angell conjectured arguing within 
the entropy model \cite{2ang95}.

The shoving model basically involves three postulates:
1) The main contribution to the activation energy is {\it 
elastic energy}; 2) This elastic energy is located in the 
{\it surroundings} of the reorienting molecules; 3) The elastic
energy is mainly
shear energy.
It is interesting to note that the model of F. Bueche from 1959
\cite{bue59} also focusses on the elastic properties of the
surroundings:
A particular molecule is regarded as surrounded by spherical
shells of
molecules, shells that are bound elastically to each other.
Bueche's idea is now that if all concentric shells should vibrate
outward in
phase, the innermost shell would expand greatly, leaving the
central molecule
in a rather large hole so it could move to a new position.
To calculate the probability of this happening Bueche made some
further assumptions, leading to an expression that at high
temperatures gives a simple Arrhenius expression but at low
temperatures a VFT-expression.

Returning to the shoving model, even if one basically accepts the
above three
postulates, there are a number of points potentially leading to
deviations from
Eq.\ (\ref{gt24}):
1)  Eq.\ (\ref{gt24}) is based on a continuum approximation that
may not be
applicable on the molecular level;
2)  The ``inner'' contribution to the activation energy has been
ignored;
3)  In real flow events spherical symmetry is probably violated
to some degree,
leading to some compression of the surroundings and thus a
contribution to the
activation energy proportional to $K_\infty$;
4)  In comparing Eq.\ (\ref{gt24}) to experiment we have ignored
any temperature-dependence of $V_c$.

Reference \cite{dyr96} gave a discussion of models 
related to the shoving model.
To the best of the author's knowledge, the first to predict an
expression equivalent to Eq.\ (\ref{gt24}) was Nemilov
\cite{nem68} who - arguing quite differently - in 1968 
arrived at this expression with our $V_c$ identified
with the total region volume.
At the present meeting Buchenau presented a model also leading 
to Eq. (\ref{gt24}) for the average relaxation time \cite{buc97}.
As emphasized by Buchenau both here and previously \cite{buc92},
models of this type link short time dynamics with long time
dynamics.
At first sight such a link may seem surprising but it makes sense

because the transition itself is a very fast process
\cite{hal87}.
Finally, we note \cite{gol85} that the present mechanism may
possibly be
applied also to explain the non-Arrhenius relaxation times of
plastic crystals \cite{sug74} and orientational glasses
\cite{hoc90}.

\acknowledgements
The author wishes to thank Prof. G. Williams for drawing
attention to the work
of F. Bueche.
 This work was supported by the Danish Natural Science Research
Council.

\end{document}